\documentclass[a4paper,fleqn]{cas-dc}

\usepackage[numbers]{natbib}

\def\tsc#1{\csdef{#1}{\textsc{\lowercase{#1}}\xspace}}
\tsc{WGM}
\tsc{QE}
\tsc{EP}
\tsc{kMS}
\tsc{BEC}
\tsc{DE}

\begin{document}
\let\WriteBookmarks\relax
\def\floatpagepagefraction{1}
\def\textpagefraction{.001}
\shorttitle{}
\shortauthors{}

\title[mode = title]{Enhancement of thermal spin pumping by orbital angular momentum of rare earth iron garnet}
\tnotemark[1]

\tnotetext[1]{This document is supported by Universitas Indonesia via Grant PUTI No. NKB-995/UN2.RST/HKP.05.00/2020
}

\author[1]{Adam B. Cahaya}[
orcid=0000-0002-2068-9613]
\cormark[1]
\fnmark[1]
\ead{adam@sci.ui.ac.id}

\credit{The author confirms sole responsibility for the following: study conception and design, data collection, analysis and interpretation of results, and manuscript preparation.}

\address[1]{Department of Physics, Faculty of Mathematics and Natural Sciences, Universitas Indonesia, Depok 16424, Indonesia
}

\begin{abstract}
In a bilayer of ferromagnetic and non-magnetic metal, spin pumping can be generated by a thermal gradient. The spin current generation depends on the spin mixing conductance of the interface and the magnetic properties of the ferromagnetic layer. Due to its low intrinsic damping, rare earth iron garnet is often used for the ferromagnetic layer in the spin Seebeck experiment. However, it is actually a ferrimagnetic with antiferromagnetically coupled magnetic lattices and the contribution of rare earth magnetic lattice of rare earth iron garnet on thermal spin pumping is not well understood. Here we focus on the effect of magnetic properties of lanthanide and show that the orbital angular momentum of rare earth iron garnet enhances thermal spin current generation of lanthanide substituted yttrium iron garnet.
\end{abstract}





\maketitle

\begin{small}
\section{Introduction}

The increased Ohmic losses that are associated with decreasing size of integrated circuits is expected to the breakdown of Moore's law \cite{Marrows2011}. By employing spin degree of freedom to existing technologies, such as thermoelectric device, the manipulation of spin current alongside electric current can lead to a better efficiency and delay the breakdown\cite{PhysRevB.93.224509}. The generation of a pure spin current with no charge current also widely research because it is expected to have less Ohmic loss \cite{Bakaul2012}. At a magnetic interface, a spin current can be generated by electric field \cite{RevModPhys.87.1213, Takahashi_2008}, ferromagnetic resonance \cite{Mizukami_2001,PhysRevLett.87.217204} or thermal gradient \cite{Uchida2008,7111309}.

One of the mechanism of the spin current generation at an interface of ferromagnet and non-magnetic metal is spin Seebeck effect \cite{Bauer2012}. The magnitude of the spin current\cite{PhysRevB.81.214418}  
\begin{eqnarray}
J_s=\frac{\hbar \gamma g^{\uparrow\downarrow}}{2\pi M_sV_c}\Delta T \label{Eq.SSE}
\end{eqnarray}
is proportional to temperature different $\Delta T$ across the interface, spin mixing conductance $g^{\uparrow\downarrow}$ of the interface \cite{PhysRevLett.111.176601}, gyromagnetic ratio $\gamma$ and the inverse of the saturated magnetization $M_s$ of the ferromagnetic layer. $V_c$ is the magnetic coherence volume of the magnetic layer \cite{PhysRevB.82.099904}. At the ferromagnet side, the spin angular momentum loss can be detected as enhancement in the magnetic damping.  At the non-magnetic metal side, the spin current can be detected as an electric voltage by means of the inverse spin Hall effect \cite{Takahashi_2008}. The direction of the detected electric current is perpendicular to both spin current direction and magnetization of the ferromagnet. 

Thermal spin pumping can also occurs when the magnetic layer is antiferromagnetic, such as MnFe$_2$\cite{PhysRevLett.116.097204,PhysRevB.93.014425} or a ferrimagnetic insulator such as Y$_3$Fe$_5$O$_{12}$ (YIG) \cite{Uchida2010,Uchida2010aip}. Furthermore, transition near the magnetization compensation point of Gd$_3$Fe$_5$O$_{12}$ is observed as the switch of the polarization of the generated spin current \cite{Geprags2016,PhysRevB.99.024417}. Recently, it has been observed that spin Seebeck effect can be enhanced by substituting yttrium in ferrimagnetic YIG by rare earths \cite{ortiz2021ultrafast,Iwasaki2019}. While machine-learning has been employed for optimizing the spin Seebeck effect\cite{Iwasaki2019}, a deeper physical understanding on the role of lanthanide on thermal spin pumping is required.

In this article, we investigate the role of lanthanide magnetic moment in the thermal spin pumping of lanthanide substituted YIG, RY$_2$Fe$_5$O$_{12}$. Sec.~\ref{Sec:SSE} describes thermal spin pumping from ferrimagnet with two magnetic sublattice. Sec.~\ref{Sec:LL} discuss the coupling of Fe and Lanthanide (R) magnetic lattices and its effect on $V_c$, $M_s$ and $\gamma$. While $M_s$ and $\gamma$ of rare earth iron garnet is widely studied, their combined relation to thermal spin pumping are not so well described. We analyze the effect of R angular momentum on $g^{\uparrow\downarrow}$ and the resultant spin current in Sec.~\ref{Sec:Result} and summarize the thermal spin pumping of RY$_2$Fe$_5$O$_{12}\vert  $Pt in Sec.~\ref{Sec:Conclusion}.

\section{Thermal spin pumping from ferrimagnet}
\label{Sec:SSE}
Ref.~\cite{PhysRevB.81.214418} derived Eq.~\ref{Eq.SSE} for describing spin Seebeck effect from ferromagnet to non-magnetic metal. Thermally generated spin current arises from spin pumping from ferromagnetic layer to non-magnetic layer \cite{PhysRevB.66.224403}
\begin{equation}
\textbf{J}_{sp}=g^{\uparrow\downarrow}\textbf{m}\times\dot{\textbf{m}},
\end{equation}
and backflow spin current due to Johnson Nyquist thermal noise.
\begin{equation}
\textbf{J}_{tn}=M_sV_c\hat{\textbf{m}}\times \textbf{h}.
\end{equation}
Here $\textbf{m}$ is magnetization direction, $\textbf{h}$ is zero-averaged magnetic field noise \cite{PhysRevLett.95.016601}
\begin{equation}
\left<h_i(t)h_j(t)\right>=\frac{2k_BT\alpha}{\gamma M_sV_c}\delta_{ij}\delta(t-t'),
\end{equation}
$\alpha$ is damping constant. At the magnetic interface $\alpha$ dominantly arise from the spin mixing at the interface and the averaged spin current is proportional to the temperature difference across the interface\cite{PhysRevB.81.214418}
\begin{eqnarray}
J_s=\left<g^{\uparrow\downarrow}\textbf{m}\times\dot{\textbf{m}} + M_sV_c \hat{\textbf{m}}\times \textbf{h}\right>=\frac{\hbar\gamma g^{\uparrow\downarrow}}{2\pi M_sV_c}\Delta T.
\end{eqnarray}

\begin{figure}
\includegraphics[width=\columnwidth]{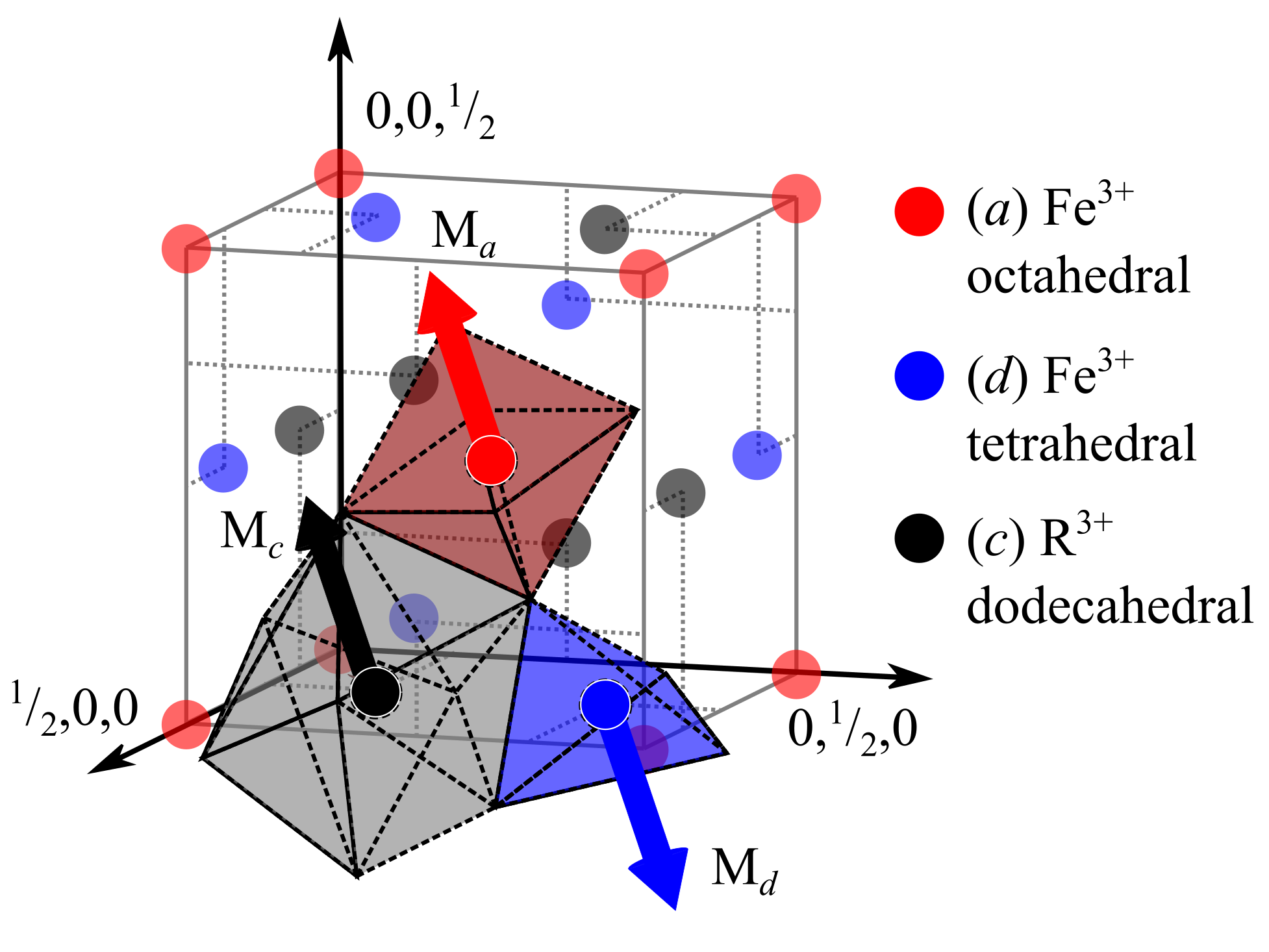}
\caption{One eighth of a unit cell of R$_3$Fe$_5$O$_{12}$ \cite{Jain2013,Vesta}. The garnet structure has metal cations (Fe$^{3+}$ and R$^{3+}$) and oxygen anions that form tetrahedral $(a)$, octahedral $(d)$ and dodecahedral $(c)$ sites \cite{GILLEO19801}. $a$ and $d$ sites are occupied by Fe$^{3+}$. $c$-site is occupied by Y$^{3+}$ or trivalent lanthanide elements (denoted by R$^{3+}$). $c$ and $a$ sites are coupled antiferromagnetically with $d$-site.
\label{Fig.RIG}}
\end{figure}

While the spin pumping of Y$_3$Fe$_5$O$_{12}$ can be described by the dynamic of Fe magnetic moment alone \cite{PhysRevB.96.144434},  the total spin current generation of RY$_2$Fe$_5$O$_{12}$ require two magnetic sub-lattices model. Since the spin mixing conductance is shown to arise from the exchange interaction between the localized spin of ferromagnet and conduction spin of the non-magnetic metal \cite{PhysRevB.96.144434,Cahaya2021,CAHAYA2022168874}
\begin{equation} 
g^{\uparrow\downarrow}\propto J_\mathrm{  ex}^2 \left(S_\mathrm{  Fe}-\vert  g_J-1\vert  J_\mathrm{  R}\right)^2, \label{Eq.spinmixing}
\end{equation} 
where $S_\mathrm{  Fe}$ and $\vert  g_J-1\vert  J_\mathrm{  R}$ are spin of iron and lanthanide (R) magnetic sublattices, respectively. $J_\mathrm{  ex}$ is the exchange constant between the localized spin of rare earth iron garnet and conduction spin of the non-magnetic metal, which mainly depend on the non-magnetic metal when the ferromagnet is an insulator, such as rare earth iron garnets \cite{PhysRevB.103.094420}.

\section{Coupled dynamics of rare earth iron garnet}
\label{Sec:LL}

The magnetization of rare earth garnet arises from the magnetic moments of trivalent ions of iron and lanthanide (R). As seen in Fig.~\ref{Fig.RIG}, Fe$^{3+}$ occupies tetrahedral ($d$) and octahedral ($a$) site, while R$^{3+}$ occupy dodecahedral ($c$) site \cite{GILLEO19801}. Since $a$ and $c$ sites are antiferromagnetically coupled to $d$-site, the total magnetization of RY$_2$Fe$_5$O$_{12}$ are 
\begin{eqnarray}
    M_s
    = M_\mathrm{  Fe}-M_\mathrm{  R}.
\end{eqnarray}
For $x=1$, the magnetization of RY$_{2}$Fe$_5$O$_{12}$ can be determined using two sub lattice model of Fe and R \cite{PhysRev.124.311}. 

While demagnetizing fields and crystallographic anisotropy may affect spin current generations, in this article we will focus on the effect of the exchange coupling of magnetic lattices. The exchange coupling can be written using interaction Hamiltonian consists of ferromagnetic coupling between Fe spins of neighboring lattice and antiferromagnetic coupling between R and Fe spins at the same lattice
\begin{eqnarray}
H_\mathrm{  int}=\lambda\sum_n \vert  g_J-1\vert  \mathbf{J}_\mathrm{  R}^{n}\cdot \mathbf{S}_\mathrm{  Fe}^{n} - I\sum_n \mathbf{S}_\mathrm{  Fe}^{n}\cdot\mathbf{S}_\mathrm{  Fe}^{n+1}.
\end{eqnarray}
Here $\mathbf{S}_\mathrm{  Fe}^{n}=\mathbf{M}_{Fe}/(2\gamma_0)$ and $\mathbf{J}_\mathrm{  R}^{(n)}=M_\mathrm{  R}/(g_J\gamma_0)$ are angular momentums of Fe and R at $n$-th site, respectively. $g_J$ is Lande $g$-factor. $I$ and $\lambda$ are the exchange constants that depends on the distances between spins. 

The Landau--Lifshitz equations of the spins under magnetic field $\mathbf{H}$ are
\begin{eqnarray}
\frac{d\mathbf{S}^{n}_{\rm Fe}}{dt}&=& \mathbf{S}^{n}_\mathrm{  Fe}\times \left(2\gamma_0\mathbf{H} + I\sum_{m=n\pm 1}\mathbf{S}^{m}_\mathrm{  Fe}
-\vert  g_J-1\vert  \mathbf{J}_\mathrm{  R}\right),\nonumber\\
\frac{d\mathbf{J}^{n}_{\rm R}}{dt}&=& \mathbf{J}^n_\mathrm{  R}\times \left(g_J\gamma_0\mathbf{H} -\vert  g_J-1\vert  \lambda \mathbf{S}^{n}_\mathrm{  Fe}\right).
\end{eqnarray}
where $\gamma_\mathrm{  0}$ is the classical gyromagnetic ratio. 
The coupled equations can be linearized by setting $F^{n}_\pm=F^{n}_x\pm iF^{n}_y=F^{\pm}e^{i(n\mathbf{k}\cdot\mathbf{a}-\omega t)}$.
\begin{equation}
\frac{d}{dt}\left[
\begin{array}{c}
S_\mathrm{  Fe}^+\\
J_\mathrm{  R}^+
\end{array}
\right]=i
W\left[
\begin{array}{c}
S_\mathrm{  Fe}^+\\
J_\mathrm{  R}^+
\end{array}
\right],
\end{equation}
\begin{small}
$$
W=\left[
\begin{array}{cc}
2\gamma_0H-4IS_\mathrm{  Fe}\sin^2\frac{ka}{2}+\vert  g_J-1\vert  \lambda J_\mathrm{  R} & \vert  g_J-1\vert  \lambda S_\mathrm{  Fe}\\
-\vert  g_J-1\vert  \lambda J_\mathrm{  R} & g_J\gamma_0H-\vert  g_J-1\vert  \lambda S_\mathrm{  Fe} 
\end{array}
\right] .
$$
\end{small}
The eigenvalues of the matrix is
\begin{eqnarray}
&\omega_{1,2}=\frac{4I\sin^2\frac{ka}{2}-\lambda'(S_\mathrm{  Fe}-J_\mathrm{  R})+(S_\mathrm{  Fe}-J_\mathrm{  R})H}{2}
\Bigg(1\pm\nonumber\\
&\sqrt{1+4\frac{\left(\lambda' S_\mathrm{  Fe}-g_J\gamma_0 H\right)\left(4I \sin^2\frac{ka}{2}+\lambda' J_\mathrm{  R}+2\gamma_0 H\right)+4\lambda'^2J_\mathrm{  R}S_\mathrm{  Fe}}{\left(4I\sin^2\frac{ka}{2}-\lambda'(S_\mathrm{  Fe}-J_\mathrm{  R})+(S_\mathrm{  Fe}-J_\mathrm{  R})H\right)^2}}\Bigg),\label{Eq.Linear}
\end{eqnarray}
where $\lambda'=\vert  g_J-1\vert  \lambda$. here $a$ and $k$ are lattice constant and wave vector. Two limits, $k=0$ ($H\neq 0$) and $H=0$ ($k\neq 0$) are useful for analyzing the effect of rare earth magnetic sub-lattice on $V_c$, $M_s$ and $\gamma$.

\subsection{Magnetic moment of lanthanide sublattice}
The spin wave modes in Fig.~\ref{Fig.spinwave} can be found from eigenfrequencies of matrix in Eq.~\ref{Eq.Linear} for $H=0$,
\begin{eqnarray}
\omega_{1}(H=0)&=& 4IS_\mathrm{  Fe} \sin^2 \frac{ka}{2}+\vert  g_J-1\vert  \lambda J_\mathrm{  R}  \label{Eq.omega1}\\
\omega_{2}(H=0)&=& -\vert  g_J-1\vert  \lambda S_\mathrm{  Fe} \label{Eq.omega2}
\end{eqnarray}
for $IS_\mathrm{  Fe}\gg \lambda (S_\mathrm{  Fe}-J_\mathrm{  R})$. 
Eq.~\ref{Eq.omega1} indicates that the changes of the spin wave stiffness 
\begin{eqnarray}
D=\lim_{ka\ll 1}\frac{1}{2}\frac{\partial^2\omega_1}{\partial k^2}\simeq IS_\mathrm{  Fe}a^2
\end{eqnarray}
due to rare earth sublattice only depends on the change of lattice constant. The effects of spin wave on the spin current generation is widely researched \cite{PhysRevB.92.054436,PhysRevB.97.020408,PhysRevLett.117.207203} and the effect of the spin wave on thermal spin pumping is captured on magnetic coherence volume $V_c$, which is directly related to the spin wave stiffness \cite{PhysRevB.82.099904}
\begin{eqnarray}
V_c\propto D^{3/2}\propto a^3.
\end{eqnarray}
 While lattice constants of R$_3$Fe$_5$O$_{12}$ is well studied \cite{PhysRevB.95.024434}, in this article we will assume that the changes of $a$ due to small R substitution RY$_2$Fe$_5$O$_{12}$, which is less than 1\% \cite{doi:10.1063/1.1733008},  is negligible.

\begin{figure}
\includegraphics[width=\columnwidth]{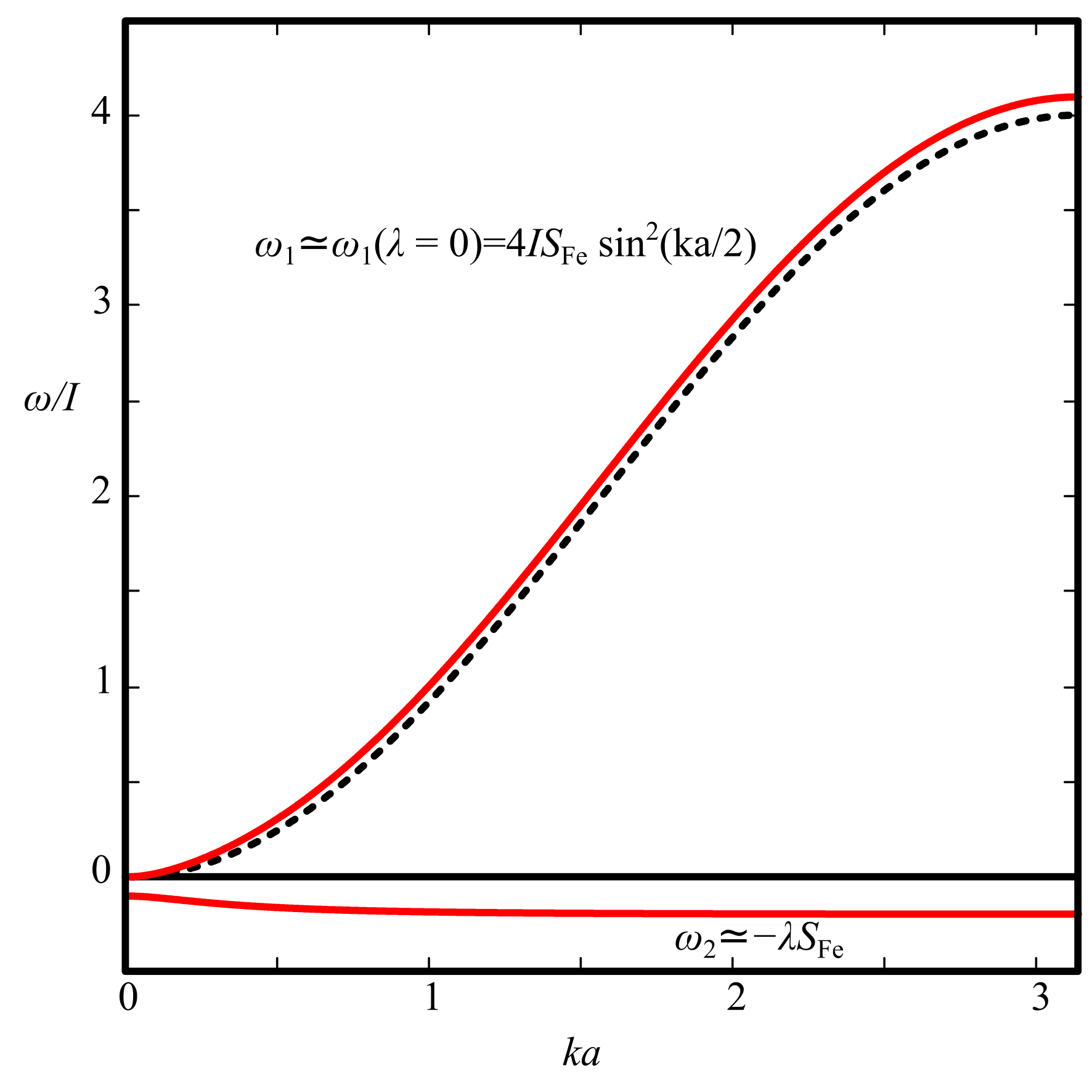}
\caption{Spin-wave spectra of two-sublattice models
of RY$_2$Fe$_5$O$_{12}$ (red lines). The spectrum of ferromagnetic spin wave $\omega_1$ of iron chain with spin $S_{\rm Fe}$ is approximately the same as the unperturbed one (black dashed line). The antiferromagnetic coupling between Fe and R generate an optical branch $\omega_2=-\lambda S_{\rm Fe}$, $\lambda$ is exchange constant of Fe and R.
\label{Fig.spinwave}}
\end{figure}

Eq.~\ref{Eq.omega2} indicates that rare earth magnetic sub-lattice of rare earth iron garnet gives an optical branch on its spin wave dispersion \cite{PhysRev.124.311}. 
The optical spin wave mode $\omega_2$ can be associated with the paramagnetic response of R to Fe molecular field. 
The temperature dependence of $M_\mathrm{  R}=g_JJ_R$ with quantum number of total angular momentum $J$ can be determined from the statical averaged of $J_R$ 
\begin{eqnarray}
    J_\mathrm{  R}(T)&=&\left(\sum_{J_z} J_ze^{-\omega_2J_z/k_BT}\right) \bigg/\left(\sum_{J_z} e^{-\omega_2J_z/k_BT}\right)\nonumber\\
    &=& JB_J\left(\frac{\vert  g_J-1\vert  \lambda{S}_\mathrm{  Fe}}{k_BT}\right),
\end{eqnarray}
$B_J$ is the Brillouin function. Since $\lim_{x\to 0} B_J(x)=(J+1)x/3$, $M_\mathrm{  R}$ at a high temperature can be approximated using Curie law
\begin{eqnarray}
    J_\mathrm{  R}(T)=\frac{\vert  g_J-1\vert  J(J+1) \lambda {S}_\mathrm{  Fe}}{3k_BT}.
    \label{Eq.Para}
\end{eqnarray}
The sign switching of $M_s=2S_\mathrm{  Fe}-xg_JJ_\mathrm{  R}$ happen at compensation point \cite{GILLEO19801,PhysRev.118.1490} 
\begin{eqnarray}
T_\mathrm{  comp}=\frac{x\lambda g_J\vert  g_J-1\vert  J(J+1)}{6k_B}. \label{Eq.Tc}
\end{eqnarray}
The sign switching of $M$ near compensation point is observed as a switching of the spin polarization of the spin current \cite{Geprags2016,PhysRevB.99.024417}.  Fig.~\ref{Fig.Tc} illustrate the relation in Eq.~\ref{Eq.Tc} for 
\begin{equation}
\lambda=
\left\{
\begin{array}{rl}
\lambda_1=&2.78 \times 10^{-2} k_BT_C, \\
\lambda_2=&\left(8.2 + 0.7 g_J(g_J-1)J(J+1) \right) \times 10^{-2} k_BT_C, 
\end{array}
\right.
\end{equation}
obtained from linear and parabolic fitting of $T_{\rm comp}$ to experimental data from Ref.~\cite{GILLEO19801}.

\begin{figure}
\includegraphics[width=\columnwidth]{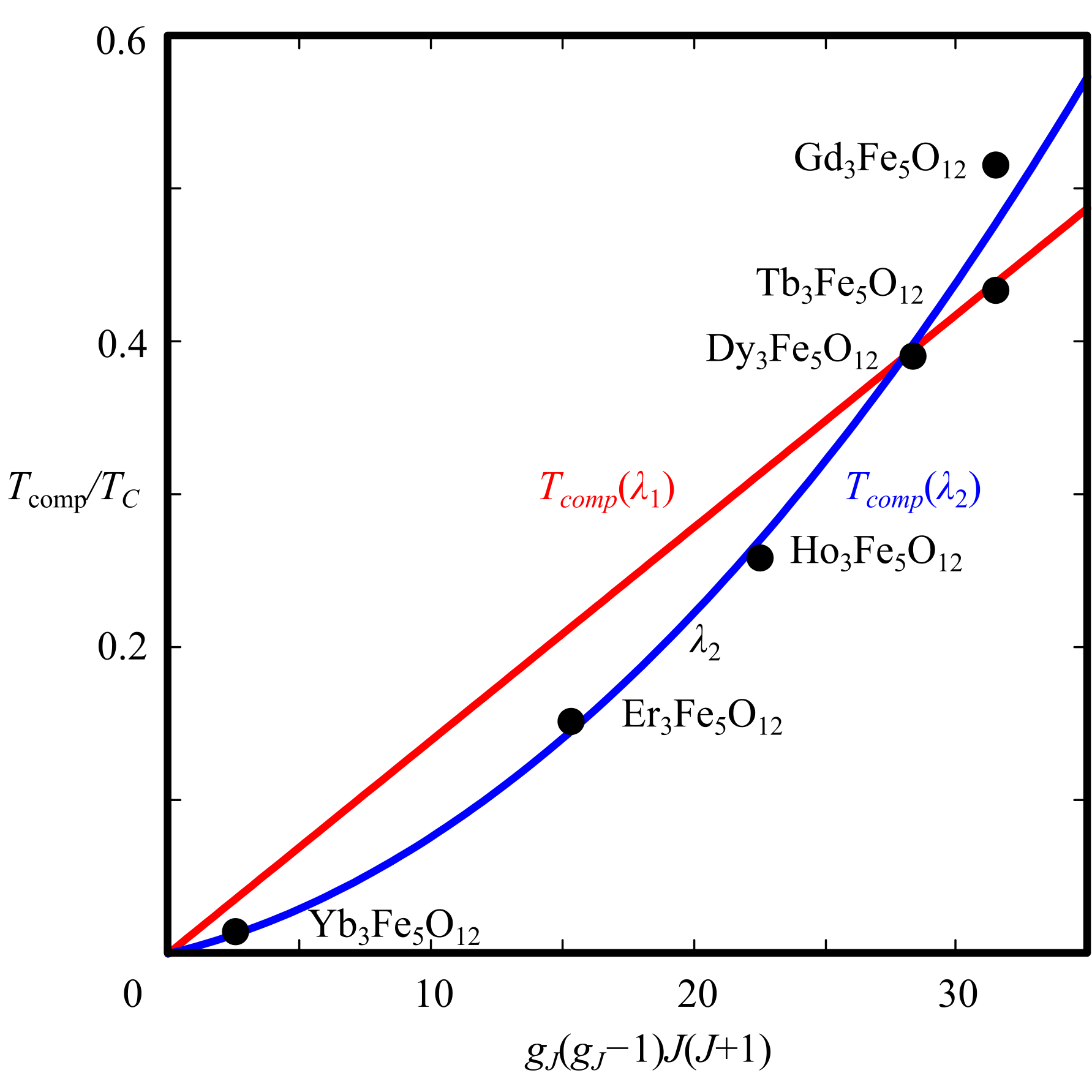}
\caption{Compensation temperature $T_\mathrm{  comp}$ of R$_3$Fe$_5$O$_{12}$ is proportional to $\lambda g_J\vert  g_J-1\vert  J(J+1)$. The red line corresponds to $\lambda_1=2.78\times 10^{-2}k_BT_C$, obtained from linear fitting to experiment data from Ref.~\cite{GILLEO19801}.
The blue lines corresponds to $\lambda_2=\left(8.2+0.7 g_J\vert  g_J-1\vert  J(J+1)\right)\times 10^{-3}k_BT_C$ obtained from parabolic fitting for better agreement. $T_C$ is the Curie temperature. At the compensation temperature total magnetization $M_s=M_\mathrm{  Fe}-M_\mathrm{  R}$ changes sign.
\label{Fig.Tc}}
\end{figure}

\subsection{Gyromagnetic ratio}
Gyromagnetic ratio $\gamma$ of RY$_2$Fe$_5$O$_{12}$ can be determined by considering $k=0$ with  $\mathbf{H}\neq 0$ limit of Eq.~\ref{Eq.Linear}
\begin{eqnarray}
\omega_{1}(k=0)&=&\vert  g_J-1\vert  \lambda\left(J_\mathrm{  R}-S_\mathrm{  Fe}\right)+\mathcal{O}(H),\label{Eq.wH1}\\
\omega_{2}(k=0)&=&\gamma_0 H\frac{2S_\mathrm{  Fe}-g_JJ_\mathrm{  R}}{S_\mathrm{  Fe}-J_\mathrm{  R}}+\mathcal{O}(H^2).\label{Eq.wH2}
\end{eqnarray}
Eq.~\ref{Eq.wH1} is related to Kaplan--Kittel frequency of two-sublattice system \cite{PhysRevB.95.024434,doi:10.1063/1.1699018} in THz regime \cite{PhysRevB.102.174432}. The resultant gyromagnetic ratio can be obtained from the lower eigenvalue (Eq.~\ref{Eq.wH2})
\begin{eqnarray}
    \gamma=\left[\frac{\partial \omega_2(k=0)}{\partial H}\right]_{H=0}=\gamma_0\frac{2S_\mathrm{  Fe}-g_JJ_\mathrm{  R}}{S_\mathrm{  Fe}-J_\mathrm{  R}}.
\end{eqnarray}
Since orbital angular momentum is $L=(2-g_J)J$, one can see orbital angular momentum increase $\gamma$ of the rare earth element
\begin{eqnarray}
    \lim_{J_\mathrm{  R}\ll S_\mathrm{  Fe}}\gamma\approx& 2\gamma_{0}\left(1+\frac{2-g_J}{2}\frac{J_\mathrm{  R}}{S_\mathrm{  Fe}}\right)= 2\gamma_{0} \left(1+\frac{L_R}{2S_\mathrm{  Fe}}\right) \label{Eq.gyroL}.
\end{eqnarray}

\section{Enhanced spin current generation}
\label{Sec:Result}

\begin{figure}[t]
\includegraphics[width=\columnwidth]{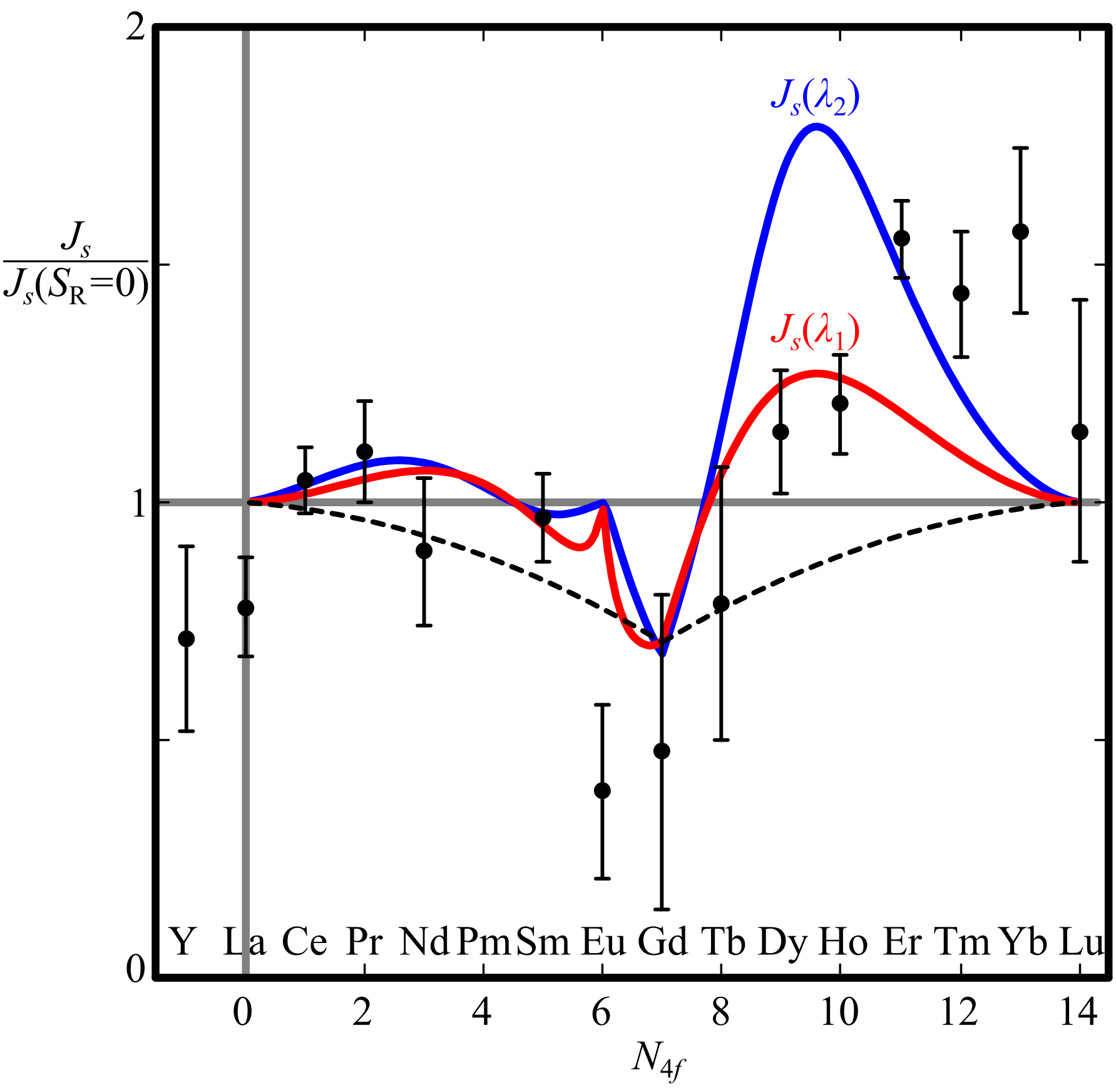}
\caption{Spin current generated by thermal spin pumping in RY$_2$Fe$_5$O$_{12}\vert  $Pt bilayer. The spin current can be enhanced by substituting one of yttriums with a lanthanide element. The experimental data was taken from spin Seebeck effect of substituted Y$_3$Fe$_5$O$_{12}\vert  $Pt from Ref.~\cite{Iwasaki2019}, tabulated in Table~\ref{Table.data}. The value is normalized to the averaged value of LaY$_2$Fe$_5$O$_{12}$ and LuY$_2$Fe$_5$O$_{12}$.
Red and blue lines show the values for exchange constant $\lambda_1$ and $\lambda_2$, respectively. The former captures the general trends of the spin current enhancement. The correction of $\lambda_2$ from parabolic fitting in Fig.~\ref{Fig.spinwave} matches the values of ErY$_2$Fe$_5$O$_{12}$ and TmY$_2$Fe$_5$O$_{12}$. Dashed black line shows the values when the spin-orbit coupling of rare earth is weak and only the spin angular momentums contributes to the magnetization and thermal spin pumping.
\label{Fig.RY2IG}}
\end{figure}

By combining the influence of lanthanide on $V_c$, $M_s$, $\gamma$ and $g^{\uparrow\downarrow}$, we can arrive at the following proportionality
\begin{eqnarray}
J_s\propto \frac{\left(S_\mathrm{  Fe}-\vert  g_J-1\vert  J_\mathrm{  R}\right)^2}{S_\mathrm{  Fe}-J_\mathrm{  R}}
\end{eqnarray}
For heavier lanthanide $g_J>1$, one can see that the spin current is enhanced by orbital angular momentum 
\begin{eqnarray}
\lim_{J_\mathrm{  R}\ll S_\mathrm{  Fe}} J_s\propto \left(1+\frac{(3-2g_J)J_\mathrm{  R}}{S_\mathrm{  Fe}}\right)=\left(1+\frac{L_\mathrm{  R}-S_\mathrm{  R}}{S_\mathrm{  Fe}}\right)
\label{Eq.total}
\end{eqnarray}
Here, $L_\mathrm{  R}$ and $S_\mathrm{  R}$ are coupled to $J_\mathrm{  R}$, which expectation value is described in Eq.~\ref{Eq.Para}.
Fig.~\ref{Fig.RY2IG} illustrates the values of thermal spin pumping of RY$_2$Fe$_5$O$_{12}\vert  $Pt compared to those of Y$_3$Fe$_5$O$_{12}\vert  $Pt in room temperature $T\sim 0.5 T_C$ and shows the agreement to experiment data by Ref.~\cite{Iwasaki2019}. 

The larger value of LuY$_2$Fe$_5$O$_{12}$ compared to LaY$_2$Fe$_5$O$_{12}$ may come from the decrease of rare earth garnet lattice constant. The decreasing value may also rise from the low heat conductance of YIG, which reduces temperature different at the interface \cite{PhysRevLett.111.176601}. By adding a corrected exchange constant ($\lambda_2$) obtained from parabolic fitting in Fig.~\ref{Fig.Tc}, the resultant thermal spin pumping agrees with the values for heavy end of lanthanide series.

\newpage
\section{Conclusion}
\label{Sec:Conclusion}

To summarize, we consider two magnetic sub-lattice model for describing the thermal spin pumping of rare earth iron garnet. The effect of lanthanide magnetic moment on the thermal spin pumping of rare earth iron garnet arise from its contribution on $M_s$, $\gamma$ and $g^{\uparrow\downarrow}$. While the angular momentum of R does not affect $V_c$ and reduces $g^{\uparrow\downarrow}$, it enhances the thermal spin pumping by reducing $M_s$ and increasing $\gamma$.

Magnetic moment of lanthanide is paramagnetically response to the molecular field of Fe. Generally, the exchange coupling constant between Fe and R can be assumed constant ($\lambda=\lambda_1$) for all lanthanide elements. A correction factor may be added from parabolic fitting in Fig.~\ref{Fig.Tc}. The resultant thermal spin pumping agrees with the values for heavy end of lanthanide series. The maximum value of thermal spin pumping is achieved when the difference between orbital and spin angular momentum of the rare earth is maximum. 

The deviation between theoretical and experimental values may arise from the dependency of crystalline and interface parameters to the lanthanide substitution, such as lattice parameter, $g$-factor and crystalline anisotropy.  The present modeling is also applicable for other ferrimagnetic materials. 

\begin{table}[t]
\centering
\caption{Voltage ($\mu$V/K) generated from spin Seebeck effect, which is proportional to the thermally generated spin current in Eq.~ \ref{Eq.SSE}. The tabulated data is extracted from Ref.~\cite{Iwasaki2019}. RY$_2$Fe$_5$O$_{12}|$Pt was grown in two kinds of substrates: Gd$_3$Ga$_5$O$_{12}$ ( GGG ) and Gd$_{2.675}$Ca$_{0.325}$Ga$_{4.025}$Mg$_{0.325}$Zr$_{0.65}$O$_{12}$ ( SGGG ). The spin Seebeck voltage of YIG is set to 2.46 $\mu$V/K.}
\label{Table.data}
\begin{tabular}{c|cc|c}
\hline &	&	 & \\[-1.5ex]
R & GGG & SGGG & weighted averaged  \\[1ex]
\hline &	&	  & 	\\[-1.5ex]
Y &	2.46$\pm$0.09 & 1.32$\pm$0.24 & 2.14$\pm$0.58 \\[1ex]
La&	2.23$\pm$0.59 & 2.36$\pm$0.10 & 2.34$\pm$0.31 \\[1ex]
Ce&	3.09$\pm$0.36 & 3.17$\pm$0.19 & 3.14$\pm$0.21 \\[1ex]
Pr&	2.89$\pm$0.18 & 3.51$\pm$0.08 & 3.32$\pm$0.32 \\[1ex]
Nd&	2.09$\pm$0.29 & 2.96$\pm$0.12 & 2.69$\pm$0.46 \\[1ex]
Sm&	3.23$\pm$0.18 & 2.71$\pm$0.10 & 2.90$\pm$0.28 \\[1ex]
Eu&	1.72$\pm$0.14 & 0.65$\pm$0.13 & 1.18$\pm$0.55 \\[1ex]
Gd&	2.70$\pm$0.15 & 0.72$\pm$0.09 & 1.43$\pm$0.99 \\[1ex]
Tb&	2.74$\pm$0.10 & 1.04$\pm$0.34 & 2.36$\pm$0.86 \\[1ex]
Dy&	3.87$\pm$0.31 & 3.17$\pm$0.19 & 3.44$\pm$0.39 \\[1ex]
Ho&	3.92$\pm$0.09 & 3.31$\pm$0.09 & 3.62$\pm$0.31 \\[1ex]
Er&	4.77$\pm$0.20 & 4.48$\pm$0.33 & 4.66$\pm$0.24 \\[1ex]
Tm&	4.54$\pm$0.12 & 3.81$\pm$0.24 & 4.31$\pm$0.39 \\[1ex]
Yb&	5.35$\pm$0.30 & 4.37$\pm$0.16 & 4.71$\pm$0.52 \\[1ex]
Lu&	3.91$\pm$0.05 & 2.27$\pm$0.13 & 3.45$\pm$0.82 \\[1ex]
\hline
\end{tabular}
\end{table}

\section*{Declaration of competing interest}
The authors declare that they have no known competing financial interests or personal relationships that could have appeared to influence the work reported in this paper.

\bibliographystyle{elsarticle-num-names}

\begin{thebibliography}{39}
\expandafter\ifx\csname natexlab\endcsname\relax\def\natexlab#1{#1}\fi
\providecommand{\url}[1]{\texttt{#1}}
\providecommand{\href}[2]{#2}
\providecommand{\path}[1]{#1}
\providecommand{\DOIprefix}{doi:}
\providecommand{\ArXivprefix}{arXiv:}
\providecommand{\URLprefix}{URL: }
\providecommand{\Pubmedprefix}{pmid:}
\providecommand{\doi}[1]{\href{http://dx.doi.org/#1}{\path{#1}}}
\providecommand{\Pubmed}[1]{\href{pmid:#1}{\path{#1}}}
\providecommand{\bibinfo}[2]{#2}
\ifx\xfnm\relax \def\xfnm[#1]{\unskip,\space#1}\fi
\bibitem[{Marrows and Hickey(2011)}]{Marrows2011}
\bibinfo{author}{C.~H. Marrows}, \bibinfo{author}{B.~J. Hickey},
\newblock \bibinfo{title}{New directions in spintronics},
\newblock \bibinfo{journal}{Philosophical Transactions of the Royal Society A:
  Mathematical, Physical and Engineering Sciences} \bibinfo{volume}{369}
  (\bibinfo{year}{2011}) \bibinfo{pages}{3027--3036}. \URLprefix
  \url{https://doi.org/10.1098/rsta.2011.0156}.
  \DOIprefix\doi{10.1098/rsta.2011.0156}.
\bibitem[{Linder and Bathen(2016)}]{PhysRevB.93.224509}
\bibinfo{author}{J.~Linder}, \bibinfo{author}{M.~E. Bathen},
\newblock \bibinfo{title}{Spin caloritronics with superconductors: Enhanced
  thermoelectric effects, generalized Onsager response-matrix, and thermal spin
  currents},
\newblock \bibinfo{journal}{Phys. Rev. B} \bibinfo{volume}{93}
  (\bibinfo{year}{2016}) \bibinfo{pages}{224509}. \URLprefix
  \url{https://link.aps.org/doi/10.1103/PhysRevB.93.224509}.
  \DOIprefix\doi{10.1103/PhysRevB.93.224509}.
\bibitem[{Bakaul et~al.(2012)Bakaul, Hu, and Kimura}]{Bakaul2012}
\bibinfo{author}{S.~R. Bakaul}, \bibinfo{author}{S.~Hu},
  \bibinfo{author}{T.~Kimura},
\newblock \bibinfo{title}{Large pure spin current generation in metallic
  nanostructures},
\newblock \bibinfo{journal}{Applied Physics A} \bibinfo{volume}{111}
  (\bibinfo{year}{2012}) \bibinfo{pages}{355--360}. \URLprefix
  \url{https://doi.org/10.1007/s00339-012-7495-0}.
  \DOIprefix\doi{10.1007/s00339-012-7495-0}.
\bibitem[{Sinova et~al.(2015)Sinova, Valenzuela, Wunderlich, Back, and
  Jungwirth}]{RevModPhys.87.1213}
\bibinfo{author}{J.~Sinova}, \bibinfo{author}{S.~O. Valenzuela},
  \bibinfo{author}{J.~Wunderlich}, \bibinfo{author}{C.~H. Back},
  \bibinfo{author}{T.~Jungwirth},
\newblock \bibinfo{title}{Spin Hall effects},
\newblock \bibinfo{journal}{Rev. Mod. Phys.} \bibinfo{volume}{87}
  (\bibinfo{year}{2015}) \bibinfo{pages}{1213--1260}. \URLprefix
  \url{https://link.aps.org/doi/10.1103/RevModPhys.87.1213}.
  \DOIprefix\doi{10.1103/RevModPhys.87.1213}.
\bibitem[{Takahashi and Maekawa(2008)}]{Takahashi_2008}
\bibinfo{author}{S.~Takahashi}, \bibinfo{author}{S.~Maekawa},
\newblock \bibinfo{title}{Spin current, spin accumulation and spin Hall
  effect},
\newblock \bibinfo{journal}{Science and Technology of Advanced Materials}
  \bibinfo{volume}{9} (\bibinfo{year}{2008}) \bibinfo{pages}{014105}.
  \URLprefix \url{https://doi.org/10.1088/1468-6996/9/1/014105}.
  \DOIprefix\doi{10.1088/1468-6996/9/1/014105}.
\bibitem[{Mizukami et~al.(2001)Mizukami, Ando, and Miyazaki}]{Mizukami_2001}
\bibinfo{author}{S.~Mizukami}, \bibinfo{author}{Y.~Ando},
  \bibinfo{author}{T.~Miyazaki},
\newblock \bibinfo{title}{The study on ferromagnetic resonance linewidth for
  {NM}/80nife/{NM} ({NM}=Cu, Ta, Pd and Pt) films},
\newblock \bibinfo{journal}{Japanese Journal of Applied Physics}
  \bibinfo{volume}{40} (\bibinfo{year}{2001}) \bibinfo{pages}{580--585}.
  \URLprefix \url{https://doi.org/10.1143/jjap.40.580}.
  \DOIprefix\doi{10.1143/jjap.40.580}.
\bibitem[{Urban et~al.(2001)Urban, Woltersdorf, and
  Heinrich}]{PhysRevLett.87.217204}
\bibinfo{author}{R.~Urban}, \bibinfo{author}{G.~Woltersdorf},
  \bibinfo{author}{B.~Heinrich},
\newblock \bibinfo{title}{Gilbert damping in single and multilayer ultrathin
  films: Role of interfaces in nonlocal spin dynamics},
\newblock \bibinfo{journal}{Phys. Rev. Lett.} \bibinfo{volume}{87}
  (\bibinfo{year}{2001}) \bibinfo{pages}{217204}. \URLprefix
  \url{https://link.aps.org/doi/10.1103/PhysRevLett.87.217204}.
  \DOIprefix\doi{10.1103/PhysRevLett.87.217204}.
\bibitem[{Uchida et~al.(2008)Uchida, Takahashi, Harii, Ieda, Koshibae, Ando,
  Maekawa, and Saitoh}]{Uchida2008}
\bibinfo{author}{K.~Uchida}, \bibinfo{author}{S.~Takahashi},
  \bibinfo{author}{K.~Harii}, \bibinfo{author}{J.~Ieda},
  \bibinfo{author}{W.~Koshibae}, \bibinfo{author}{K.~Ando},
  \bibinfo{author}{S.~Maekawa}, \bibinfo{author}{E.~Saitoh},
\newblock \bibinfo{title}{Observation of the spin Seebeck effect},
\newblock \bibinfo{journal}{Nature} \bibinfo{volume}{455}
  (\bibinfo{year}{2008}) \bibinfo{pages}{778--781}. \URLprefix
  \url{https://doi.org/10.1038/nature07321}.
  \DOIprefix\doi{10.1038/nature07321}.
\bibitem[{Cahaya et~al.(2015)Cahaya, Tretiakov, and Bauer}]{7111309}
\bibinfo{author}{A.~B. Cahaya}, \bibinfo{author}{O.~A. Tretiakov},
  \bibinfo{author}{G.~E.~W. Bauer},
\newblock \bibinfo{title}{Spin Seebeck power conversion},
\newblock \bibinfo{journal}{IEEE Transactions on Magnetics}
  \bibinfo{volume}{51} (\bibinfo{year}{2015}) \bibinfo{pages}{1--14}.
  \DOIprefix\doi{10.1109/TMAG.2015.2436362}.
\bibitem[{Bauer et~al.(2012)Bauer, Saitoh, and van Wees}]{Bauer2012}
\bibinfo{author}{G.~E.~W. Bauer}, \bibinfo{author}{E.~Saitoh},
  \bibinfo{author}{B.~J. van Wees},
\newblock \bibinfo{title}{Spin caloritronics},
\newblock \bibinfo{journal}{Nature Materials} \bibinfo{volume}{11}
  (\bibinfo{year}{2012}) \bibinfo{pages}{391--399}. \URLprefix
  \url{https://doi.org/10.1038/nmat3301}. \DOIprefix\doi{10.1038/nmat3301}.
\bibitem[{Xiao et~al.(2010)Xiao, Bauer, Uchida, Saitoh, and
  Maekawa}]{PhysRevB.81.214418}
\bibinfo{author}{J.~Xiao}, \bibinfo{author}{G.~E.~W. Bauer},
  \bibinfo{author}{K.-c. Uchida}, \bibinfo{author}{E.~Saitoh},
  \bibinfo{author}{S.~Maekawa},
\newblock \bibinfo{title}{Theory of magnon-driven spin Seebeck effect},
\newblock \bibinfo{journal}{Phys. Rev. B} \bibinfo{volume}{81}
  (\bibinfo{year}{2010}) \bibinfo{pages}{214418}. \URLprefix
  \url{https://link.aps.org/doi/10.1103/PhysRevB.81.214418}.
  \DOIprefix\doi{10.1103/PhysRevB.81.214418}.
\bibitem[{Weiler et~al.(2013)Weiler, Althammer, Schreier, Lotze, Pernpeintner,
  Meyer, Huebl, Gross, Kamra, Xiao, Chen, Jiao, Bauer, and
  Goennenwein}]{PhysRevLett.111.176601}
\bibinfo{author}{M.~Weiler}, \bibinfo{author}{M.~Althammer},
  \bibinfo{author}{M.~Schreier}, \bibinfo{author}{J.~Lotze},
  \bibinfo{author}{M.~Pernpeintner}, \bibinfo{author}{S.~Meyer},
  \bibinfo{author}{H.~Huebl}, \bibinfo{author}{R.~Gross},
  \bibinfo{author}{A.~Kamra}, \bibinfo{author}{J.~Xiao}, \bibinfo{author}{Y.-T.
  Chen}, \bibinfo{author}{H.~J. Jiao}, \bibinfo{author}{G.~E.~W. Bauer},
  \bibinfo{author}{S.~T.~B. Goennenwein},
\newblock \bibinfo{title}{Experimental test of the spin mixing interface
  conductivity concept},
\newblock \bibinfo{journal}{Phys. Rev. Lett.} \bibinfo{volume}{111}
  (\bibinfo{year}{2013}) \bibinfo{pages}{176601}. \URLprefix
  \url{https://link.aps.org/doi/10.1103/PhysRevLett.111.176601}.
  \DOIprefix\doi{10.1103/PhysRevLett.111.176601}.
\bibitem[{Xiao et~al.(2010)Xiao, Bauer, Uchida, Saitoh, and
  Maekawa}]{PhysRevB.82.099904}
\bibinfo{author}{J.~Xiao}, \bibinfo{author}{G.~E.~W. Bauer},
  \bibinfo{author}{K.-C. Uchida}, \bibinfo{author}{E.~Saitoh},
  \bibinfo{author}{S.~Maekawa},
\newblock \bibinfo{title}{Erratum: Theory of magnon-driven spin Seebeck effect
  [phys. rev. b 81, 214418 (2010)]},
\newblock \bibinfo{journal}{Phys. Rev. B} \bibinfo{volume}{82}
  (\bibinfo{year}{2010}) \bibinfo{pages}{099904(E)}. \URLprefix
  \url{https://link.aps.org/doi/10.1103/PhysRevB.82.099904}.
  \DOIprefix\doi{10.1103/PhysRevB.82.099904}.
\bibitem[{Wu et~al.(2016)Wu, Zhang, Amit, Borisov, Pearson, Jiang, Lederman,
  Hoffmann, and Bhattacharya}]{PhysRevLett.116.097204}
\bibinfo{author}{S.~M. Wu}, \bibinfo{author}{W.~Zhang},
  \bibinfo{author}{K.~Amit}, \bibinfo{author}{P.~Borisov},
  \bibinfo{author}{J.~E. Pearson}, \bibinfo{author}{J.~S. Jiang},
  \bibinfo{author}{D.~Lederman}, \bibinfo{author}{A.~Hoffmann},
  \bibinfo{author}{A.~Bhattacharya},
\newblock \bibinfo{title}{Antiferromagnetic spin Seebeck effect},
\newblock \bibinfo{journal}{Phys. Rev. Lett.} \bibinfo{volume}{116}
  (\bibinfo{year}{2016}) \bibinfo{pages}{097204(B)}. \URLprefix
  \url{https://link.aps.org/doi/10.1103/PhysRevLett.116.097204}.
  \DOIprefix\doi{10.1103/PhysRevLett.116.097204}.
\bibitem[{Rezende et~al.(2016)Rezende, Rodr\'{\i}guez-Su\'arez, and
  Azevedo}]{PhysRevB.93.014425}
\bibinfo{author}{S.~M. Rezende}, \bibinfo{author}{R.~L.
  Rodr\'{\i}guez-Su\'arez}, \bibinfo{author}{A.~Azevedo},
\newblock \bibinfo{title}{Theory of the spin Seebeck effect in
  antiferromagnets},
\newblock \bibinfo{journal}{Phys. Rev. B} \bibinfo{volume}{93}
  (\bibinfo{year}{2016}) \bibinfo{pages}{014425}. \URLprefix
  \url{https://link.aps.org/doi/10.1103/PhysRevB.93.014425}.
  \DOIprefix\doi{10.1103/PhysRevB.93.014425}.
\bibitem[{Uchida et~al.(2010)Uchida, Xiao, Adachi, Ohe, Takahashi, Ieda, Ota,
  Kajiwara, Umezawa, Kawai, Bauer, Maekawa, and Saitoh}]{Uchida2010}
\bibinfo{author}{K.~Uchida}, \bibinfo{author}{J.~Xiao},
  \bibinfo{author}{H.~Adachi}, \bibinfo{author}{J.~Ohe},
  \bibinfo{author}{S.~Takahashi}, \bibinfo{author}{J.~Ieda},
  \bibinfo{author}{T.~Ota}, \bibinfo{author}{Y.~Kajiwara},
  \bibinfo{author}{H.~Umezawa}, \bibinfo{author}{H.~Kawai},
  \bibinfo{author}{G.~E.~W. Bauer}, \bibinfo{author}{S.~Maekawa},
  \bibinfo{author}{E.~Saitoh},
\newblock \bibinfo{title}{Spin Seebeck insulator},
\newblock \bibinfo{journal}{Nature Materials} \bibinfo{volume}{9}
  (\bibinfo{year}{2010}) \bibinfo{pages}{894--897}. \URLprefix
  \url{https://doi.org/10.1038/nmat2856}. \DOIprefix\doi{10.1038/nmat2856}.
\bibitem[{ichi Uchida et~al.(2010)ichi Uchida, Adachi, Ota, Nakayama, Maekawa,
  and Saitoh}]{Uchida2010aip}
\bibinfo{author}{K.~ichi Uchida}, \bibinfo{author}{H.~Adachi},
  \bibinfo{author}{T.~Ota}, \bibinfo{author}{H.~Nakayama},
  \bibinfo{author}{S.~Maekawa}, \bibinfo{author}{E.~Saitoh},
\newblock \bibinfo{title}{Observation of longitudinal spin-Seebeck effect in
  magnetic insulators},
\newblock \bibinfo{journal}{Applied Physics Letters} \bibinfo{volume}{97}
  (\bibinfo{year}{2010}) \bibinfo{pages}{172505}. \URLprefix
  \url{https://doi.org/10.1063/1.3507386}. \DOIprefix\doi{10.1063/1.3507386}.
\bibitem[{Gepr{\"a}gs et~al.(2016)Gepr{\"a}gs, Kehlberger, Coletta, Qiu, Guo,
  Schulz, Mix, Meyer, Kamra, Althammer, Huebl, Jakob, Ohnuma, Adachi, Barker,
  Maekawa, Bauer, Saitoh, Gross, Goennenwein, and Kl{\"a}ui}]{Geprags2016}
\bibinfo{author}{S.~Gepr{\"a}gs}, \bibinfo{author}{A.~Kehlberger},
  \bibinfo{author}{F.~D. Coletta}, \bibinfo{author}{Z.~Qiu},
  \bibinfo{author}{E.-J. Guo}, \bibinfo{author}{T.~Schulz},
  \bibinfo{author}{C.~Mix}, \bibinfo{author}{S.~Meyer},
  \bibinfo{author}{A.~Kamra}, \bibinfo{author}{M.~Althammer},
  \bibinfo{author}{H.~Huebl}, \bibinfo{author}{G.~Jakob},
  \bibinfo{author}{Y.~Ohnuma}, \bibinfo{author}{H.~Adachi},
  \bibinfo{author}{J.~Barker}, \bibinfo{author}{S.~Maekawa},
  \bibinfo{author}{G.~E.~W. Bauer}, \bibinfo{author}{E.~Saitoh},
  \bibinfo{author}{R.~Gross}, \bibinfo{author}{S.~T.~B. Goennenwein},
  \bibinfo{author}{M.~Kl{\"a}ui},
\newblock \bibinfo{title}{Origin of the spin Seebeck effect in compensated
  ferrimagnets},
\newblock \bibinfo{journal}{Nature Communications} \bibinfo{volume}{7}
  (\bibinfo{year}{2016}) \bibinfo{pages}{10452}. \URLprefix
  \url{https://doi.org/10.1038/ncomms10452}.
  \DOIprefix\doi{10.1038/ncomms10452}.
\bibitem[{Shen(2019)}]{PhysRevB.99.024417}
\bibinfo{author}{K.~Shen},
\newblock \bibinfo{title}{Temperature-switched anomaly in the spin Seebeck
  effect in ${\mathrm{gd}}_{3}{\mathrm{fe}}_{5}{\mathrm{o}}_{12}$},
\newblock \bibinfo{journal}{Phys. Rev. B} \bibinfo{volume}{99}
  (\bibinfo{year}{2019}) \bibinfo{pages}{024417}. \URLprefix
  \url{https://link.aps.org/doi/10.1103/PhysRevB.99.024417}.
  \DOIprefix\doi{10.1103/PhysRevB.99.024417}.
\bibitem[{Ortiz et~al.(2021)Ortiz, Gomez, Liu, Aldosary, Shi, and
  Wilson}]{ortiz2021ultrafast}
\bibinfo{author}{V.~H. Ortiz}, \bibinfo{author}{M.~J. Gomez},
  \bibinfo{author}{Y.~Liu}, \bibinfo{author}{M.~Aldosary},
  \bibinfo{author}{J.~Shi}, \bibinfo{author}{R.~B. Wilson},
\newblock \bibinfo{title}{Ultrafast measurements of the interfacial spin
  Seebeck effect in au and rare-earth iron-garnet bilayers},
\newblock \bibinfo{journal}{Phys. Rev. Materials} \bibinfo{volume}{5}
  (\bibinfo{year}{2021}) \bibinfo{pages}{074401}. \URLprefix
  \url{https://link.aps.org/doi/10.1103/PhysRevMaterials.5.074401}.
  \DOIprefix\doi{10.1103/PhysRevMaterials.5.074401}.
\bibitem[{Iwasaki et~al.(2019)Iwasaki, Takeuchi, Stanev, Kusne, Ishida,
  Kirihara, Ihara, Sawada, Terashima, Someya, Uchida, Saitoh, and
  Yorozu}]{Iwasaki2019}
\bibinfo{author}{Y.~Iwasaki}, \bibinfo{author}{I.~Takeuchi},
  \bibinfo{author}{V.~Stanev}, \bibinfo{author}{A.~G. Kusne},
  \bibinfo{author}{M.~Ishida}, \bibinfo{author}{A.~Kirihara},
  \bibinfo{author}{K.~Ihara}, \bibinfo{author}{R.~Sawada},
  \bibinfo{author}{K.~Terashima}, \bibinfo{author}{H.~Someya},
  \bibinfo{author}{K.-i. Uchida}, \bibinfo{author}{E.~Saitoh},
  \bibinfo{author}{S.~Yorozu},
\newblock \bibinfo{title}{Machine-learning guided discovery of a new
  thermoelectric material},
\newblock \bibinfo{journal}{Scientific Reports} \bibinfo{volume}{9}
  (\bibinfo{year}{2019}) \bibinfo{pages}{2751}. \URLprefix
  \url{https://doi.org/10.1038/s41598-019-39278-z}.
  \DOIprefix\doi{10.1038/s41598-019-39278-z}.
\bibitem[{Jain et~al.(2013)Jain, Ong, Hautier, Chen, Richards, Dacek, Cholia,
  Gunter, Skinner, Ceder, and Persson}]{Jain2013}
\bibinfo{author}{A.~Jain}, \bibinfo{author}{S.~P. Ong},
  \bibinfo{author}{G.~Hautier}, \bibinfo{author}{W.~Chen},
  \bibinfo{author}{W.~D. Richards}, \bibinfo{author}{S.~Dacek},
  \bibinfo{author}{S.~Cholia}, \bibinfo{author}{D.~Gunter},
  \bibinfo{author}{D.~Skinner}, \bibinfo{author}{G.~Ceder},
  \bibinfo{author}{K.~a. Persson},
\newblock \bibinfo{title}{{The Materials Project: A materials genome approach
  to accelerating materials innovation}},
\newblock \bibinfo{journal}{APL Materials} \bibinfo{volume}{1}
  (\bibinfo{year}{2013}) \bibinfo{pages}{011002}. \URLprefix
  \url{http://link.aip.org/link/AMPADS/v1/i1/p011002/s1\&Agg=doi}.
  \DOIprefix\doi{10.1063/1.4812323}.
\bibitem[{Momma and Izumi(2011)}]{Vesta}
\bibinfo{author}{K.~Momma}, \bibinfo{author}{F.~Izumi},
\newblock \bibinfo{title}{{{\it VESTA3} for three-dimensional visualization of
  crystal, volumetric and morphology data}},
\newblock \bibinfo{journal}{Journal of Applied Crystallography}
  \bibinfo{volume}{44} (\bibinfo{year}{2011}) \bibinfo{pages}{1272--1276}.
  \URLprefix \url{https://doi.org/10.1107/S0021889811038970}.
  \DOIprefix\doi{10.1107/S0021889811038970}.
\bibitem[{Gilleo(1980)}]{GILLEO19801}
\bibinfo{author}{M.~Gilleo},
\newblock \bibinfo{title}{Chapter 1 ferromagnetic insulators: Garnets},
\newblock volume~\bibinfo{volume}{2} of \textit{\bibinfo{series}{Handbook of
  Ferromagnetic Materials}}, \bibinfo{publisher}{Elsevier},
  \bibinfo{year}{1980}, pp. \bibinfo{pages}{1--53}. \URLprefix
  \url{https://www.sciencedirect.com/science/article/pii/S1574930405801026}.
  \DOIprefix\doi{https://doi.org/10.1016/S1574-9304(05)80102-6}.
\bibitem[{Tserkovnyak et~al.(2002)Tserkovnyak, Brataas, and
  Bauer}]{PhysRevB.66.224403}
\bibinfo{author}{Y.~Tserkovnyak}, \bibinfo{author}{A.~Brataas},
  \bibinfo{author}{G.~E.~W. Bauer},
\newblock \bibinfo{title}{Spin pumping and magnetization dynamics in metallic
  multilayers},
\newblock \bibinfo{journal}{Phys. Rev. B} \bibinfo{volume}{66}
  (\bibinfo{year}{2002}) \bibinfo{pages}{224403}. \URLprefix
  \url{https://link.aps.org/doi/10.1103/PhysRevB.66.224403}.
  \DOIprefix\doi{10.1103/PhysRevB.66.224403}.
\bibitem[{Foros et~al.(2005)Foros, Brataas, Tserkovnyak, and
  Bauer}]{PhysRevLett.95.016601}
\bibinfo{author}{J.~Foros}, \bibinfo{author}{A.~Brataas},
  \bibinfo{author}{Y.~Tserkovnyak}, \bibinfo{author}{G.~E.~W. Bauer},
\newblock \bibinfo{title}{Magnetization noise in magnetoelectronic
  nanostructures},
\newblock \bibinfo{journal}{Phys. Rev. Lett.} \bibinfo{volume}{95}
  (\bibinfo{year}{2005}) \bibinfo{pages}{016601}. \URLprefix
  \url{https://link.aps.org/doi/10.1103/PhysRevLett.95.016601}.
  \DOIprefix\doi{10.1103/PhysRevLett.95.016601}.
\bibitem[{Cahaya et~al.(2017)Cahaya, Leon, and Bauer}]{PhysRevB.96.144434}
\bibinfo{author}{A.~B. Cahaya}, \bibinfo{author}{A.~O. Leon},
  \bibinfo{author}{G.~E.~W. Bauer},
\newblock \bibinfo{title}{Crystal field effects on spin pumping},
\newblock \bibinfo{journal}{Phys. Rev. B} \bibinfo{volume}{96}
  (\bibinfo{year}{2017}) \bibinfo{pages}{144434}. \URLprefix
  \url{https://link.aps.org/doi/10.1103/PhysRevB.96.144434}.
  \DOIprefix\doi{10.1103/PhysRevB.96.144434}.
\bibitem[{Cahaya(2021)}]{Cahaya2021}
\bibinfo{author}{A.~B. Cahaya},
\newblock \bibinfo{title}{Antiferromagnetic spin pumping via hyperfine
  interaction},
\newblock \bibinfo{journal}{Hyperfine Interactions} \bibinfo{volume}{242}
  (\bibinfo{year}{2021}) \bibinfo{pages}{46}. \URLprefix
  \url{https://doi.org/10.1007/s10751-021-01780-0}.
  \DOIprefix\doi{10.1007/s10751-021-01780-0}.
\bibitem[{Cahaya(2022)}]{CAHAYA2022168874}
\bibinfo{author}{A.~B. Cahaya},
\newblock \bibinfo{title}{Adiabatic limit of RKKY range function in one
  dimension},
\newblock \bibinfo{journal}{Journal of Magnetism and Magnetic Materials}
  \bibinfo{volume}{547} (\bibinfo{year}{2022}) \bibinfo{pages}{168874}.
  \URLprefix
  \url{https://www.sciencedirect.com/science/article/pii/S0304885321010805}.
  \DOIprefix\doi{https://doi.org/10.1016/j.jmmm.2021.168874}.
\bibitem[{Cahaya and Majidi(2021)}]{PhysRevB.103.094420}
\bibinfo{author}{A.~B. Cahaya}, \bibinfo{author}{M.~A. Majidi},
\newblock \bibinfo{title}{Effects of screened Coulomb interaction on spin
  transfer torque},
\newblock \bibinfo{journal}{Phys. Rev. B} \bibinfo{volume}{103}
  (\bibinfo{year}{2021}) \bibinfo{pages}{094420}. \URLprefix
  \url{https://link.aps.org/doi/10.1103/PhysRevB.103.094420}.
  \DOIprefix\doi{10.1103/PhysRevB.103.094420}.
\bibitem[{Tinkham(1961)}]{PhysRev.124.311}
\bibinfo{author}{M.~Tinkham},
\newblock \bibinfo{title}{Low-lying spectrum of rare-earth iron garnets},
\newblock \bibinfo{journal}{Phys. Rev.} \bibinfo{volume}{124}
  (\bibinfo{year}{1961}) \bibinfo{pages}{311--320}. \URLprefix
  \url{https://link.aps.org/doi/10.1103/PhysRev.124.311}.
  \DOIprefix\doi{10.1103/PhysRev.124.311}.
\bibitem[{Jin et~al.(2015)Jin, Boona, Yang, Myers, and
  Heremans}]{PhysRevB.92.054436}
\bibinfo{author}{H.~Jin}, \bibinfo{author}{S.~R. Boona},
  \bibinfo{author}{Z.~Yang}, \bibinfo{author}{R.~C. Myers},
  \bibinfo{author}{J.~P. Heremans},
\newblock \bibinfo{title}{Effect of the magnon dispersion on the longitudinal
  spin Seebeck effect in yttrium iron garnets},
\newblock \bibinfo{journal}{Phys. Rev. B} \bibinfo{volume}{92}
  (\bibinfo{year}{2015}) \bibinfo{pages}{054436}. \URLprefix
  \url{https://link.aps.org/doi/10.1103/PhysRevB.92.054436}.
  \DOIprefix\doi{10.1103/PhysRevB.92.054436}.
\bibitem[{Prakash et~al.(2018)Prakash, Flebus, Brangham, Yang, Tserkovnyak, and
  Heremans}]{PhysRevB.97.020408}
\bibinfo{author}{A.~Prakash}, \bibinfo{author}{B.~Flebus},
  \bibinfo{author}{J.~Brangham}, \bibinfo{author}{F.~Yang},
  \bibinfo{author}{Y.~Tserkovnyak}, \bibinfo{author}{J.~P. Heremans},
\newblock \bibinfo{title}{Evidence for the role of the magnon energy relaxation
  length in the spin Seebeck effect},
\newblock \bibinfo{journal}{Phys. Rev. B} \bibinfo{volume}{97}
  (\bibinfo{year}{2018}) \bibinfo{pages}{020408}. \URLprefix
  \url{https://link.aps.org/doi/10.1103/PhysRevB.97.020408}.
  \DOIprefix\doi{10.1103/PhysRevB.97.020408}.
\bibitem[{Kikkawa et~al.(2016)Kikkawa, Shen, Flebus, Duine, Uchida, Qiu, Bauer,
  and Saitoh}]{PhysRevLett.117.207203}
\bibinfo{author}{T.~Kikkawa}, \bibinfo{author}{K.~Shen},
  \bibinfo{author}{B.~Flebus}, \bibinfo{author}{R.~A. Duine},
  \bibinfo{author}{K.-i. Uchida}, \bibinfo{author}{Z.~Qiu},
  \bibinfo{author}{G.~E.~W. Bauer}, \bibinfo{author}{E.~Saitoh},
\newblock \bibinfo{title}{Magnon polarons in the spin Seebeck effect},
\newblock \bibinfo{journal}{Phys. Rev. Lett.} \bibinfo{volume}{117}
  (\bibinfo{year}{2016}) \bibinfo{pages}{207203}. \URLprefix
  \url{https://link.aps.org/doi/10.1103/PhysRevLett.117.207203}.
  \DOIprefix\doi{10.1103/PhysRevLett.117.207203}.
\bibitem[{Nakamoto et~al.(2017)Nakamoto, Xu, Xu, Xu, and
  Bellaiche}]{PhysRevB.95.024434}
\bibinfo{author}{R.~Nakamoto}, \bibinfo{author}{B.~Xu},
  \bibinfo{author}{C.~Xu}, \bibinfo{author}{H.~Xu},
  \bibinfo{author}{L.~Bellaiche},
\newblock \bibinfo{title}{Properties of rare-earth iron garnets from first
  principles},
\newblock \bibinfo{journal}{Phys. Rev. B} \bibinfo{volume}{95}
  (\bibinfo{year}{2017}) \bibinfo{pages}{024434}. \URLprefix
  \url{https://link.aps.org/doi/10.1103/PhysRevB.95.024434}.
  \DOIprefix\doi{10.1103/PhysRevB.95.024434}.
\bibitem[{Espinosa(1962)}]{doi:10.1063/1.1733008}
\bibinfo{author}{G.~P. Espinosa},
\newblock \bibinfo{title}{Crystal chemical study of the rare‐earth iron
  garnets},
\newblock \bibinfo{journal}{The Journal of Chemical Physics}
  \bibinfo{volume}{37} (\bibinfo{year}{1962}) \bibinfo{pages}{2344--2347}.
  \URLprefix \url{https://doi.org/10.1063/1.1733008}.
  \DOIprefix\doi{10.1063/1.1733008}.
  \href{http://arxiv.org/abs/https://doi.org/10.1063/1.1733008}{{\tt
  arXiv:https://doi.org/10.1063/1.1733008}}.
\bibitem[{Wolf and Van~Vleck(1960)}]{PhysRev.118.1490}
\bibinfo{author}{W.~P. Wolf}, \bibinfo{author}{J.~H. Van~Vleck},
\newblock \bibinfo{title}{Magnetism of europium garnet},
\newblock \bibinfo{journal}{Phys. Rev.} \bibinfo{volume}{118}
  (\bibinfo{year}{1960}) \bibinfo{pages}{1490--1492}. \URLprefix
  \url{https://link.aps.org/doi/10.1103/PhysRev.118.1490}.
  \DOIprefix\doi{10.1103/PhysRev.118.1490}.
\bibitem[{Kaplan and Kittel(1953)}]{doi:10.1063/1.1699018}
\bibinfo{author}{J.~Kaplan}, \bibinfo{author}{C.~Kittel},
\newblock \bibinfo{title}{Exchange frequency electron spin resonance in
  ferrites},
\newblock \bibinfo{journal}{The Journal of Chemical Physics}
  \bibinfo{volume}{21} (\bibinfo{year}{1953}) \bibinfo{pages}{760--761}.
  \URLprefix \url{https://doi.org/10.1063/1.1699018}.
  \DOIprefix\doi{10.1063/1.1699018}.
  \href{http://arxiv.org/abs/https://doi.org/10.1063/1.1699018}{{\tt
  arXiv:https://doi.org/10.1063/1.1699018}}.
\bibitem[{Hsu et~al.(2020)Hsu, Shen, Fujii, Koreeda, and
  Satoh}]{PhysRevB.102.174432}
\bibinfo{author}{W.-H. Hsu}, \bibinfo{author}{K.~Shen},
  \bibinfo{author}{Y.~Fujii}, \bibinfo{author}{A.~Koreeda},
  \bibinfo{author}{T.~Satoh},
\newblock \bibinfo{title}{Observation of terahertz magnon of Kaplan-Kittel
  exchange resonance in yttrium-iron garnet by raman spectroscopy},
\newblock \bibinfo{journal}{Phys. Rev. B} \bibinfo{volume}{102}
  (\bibinfo{year}{2020}) \bibinfo{pages}{174432}. \URLprefix
  \url{https://link.aps.org/doi/10.1103/PhysRevB.102.174432}.
  \DOIprefix\doi{10.1103/PhysRevB.102.174432}.

\end{thebibliography}

\end{small}
\end{document}